\documentclass[preprint,showpacs,preprintnumbers,amsmath,amssymb]{revtex4}

\usepackage{graphicx}
\usepackage{dcolumn}
\usepackage{bm}

\def\beq{\begin{equation}}
\def\eeq{\end{equation}}
\def\eea{\end{eqnarray}}
\def\bea{\begin{eqnarray}}

\def\tev{\, {\rm TeV}}
\def\gev{\, {\rm GeV}}
\def\swsq{\sin^2\theta_W}
\newcommand{\gsim}{\lower.7ex\hbox{$\;\stackrel{\textstyle>}{\sim}\;$}}
\newcommand{\lsim}{\lower.7ex\hbox{$\;\stackrel{\textstyle<}{\sim}\;$}}

\def\slashchar#1{\setbox0=\hbox{$#1$}           
   \dimen0=\wd0                                 
   \setbox1=\hbox{/} \dimen1=\wd1               
   \ifdim\dimen0>\dimen1                        
      \rlap{\hbox to \dimen0{\hfil/\hfil}}      
      #1                                        
   \else                                        
      \rlap{\hbox to \dimen1{\hfil$#1$\hfil}}   
      /                                         
   \fi}                                         %

\begin{document}

\preprint{hep-ph/0204196}

\title{Higgs boson mass limits in perturbative unification theories}
\author{Kazuhiro Tobe and James D. Wells}

\address{Physics Department, University of California, 
Davis, CA 95616}


\date{\today}

\begin{abstract}

Motivated in part by recent demonstrations that electroweak unification
into a simple group
may occur at a low scale, we detail the requirements on the Higgs
mass if the unification is to be perturbative.  We do this for the
Standard Model effective theory, minimal supersymmetry, and 
next-to-minimal supersymmetry with
an additional singlet field.  Within the Standard Model framework, we
find that perturbative unification with $\swsq=1/4$ occurs
at $\Lambda=3.8\tev$ and requires
$m_h\lsim 460$ GeV, whereas perturbative unification with
$\swsq=3/8$ requires $m_h \lsim 200$ GeV.
In supersymmetry, the
presentation of the Higgs mass predictions
can be significantly simplified, yet remain meaningful, by using a 
single supersymmetry breaking parameter $\Delta_S$.  We present Higgs
mass limits
in terms of $\Delta_S$ for the minimal supersymmetric model
and the next-to-minimal supersymmetric model.
We show that in next-to-minimal supersymmetry, the Higgs mass
upper limit can be as large as $500\gev$ even for moderate supersymmetry
masses if the perturbative unification scale is low ($\Lambda\simeq 10\tev$).

\end{abstract}

\pacs{14.80.Bn, 12.10.Kt, 12.60.Jv}

\maketitle

\section{Introduction}

A large gap in our understanding of fundamental physics is the mechanism
of electroweak symmetry breaking and fermion mass generation.  Among the
many ideas developed to explain this phenomena, the most economical
explanation postulates the existence of a single
scalar Higgs boson. This simple explanation
has been remarkably successful, in that all precision electroweak
data is compatible with it, yet
not compatible with many other more complicated explanations of dynamical
electroweak symmetry breaking.

Despite the success of the single Higgs boson theory, there are two
challenges.  First, there are theoretical problems explaining the large
hierarchy of fundamental scales (e.g., $M_{\rm Pl}\gg m_W$).  And second,
we have yet to find the Higgs boson in experiment.

The Standard Model (SM) Higgs properties are completely fixed in terms of
only one parameter, its mass.  Unfortunately, the mass cannot be
predicted.  Precision measurements have constrained the Higgs boson
mass to be below $222\gev$ at the 95\% C.L.\ (see p.\ 101 
of~\cite{Abbaneo:2001ix}).  
Direct searches
in $e^+e^-\to Z +{\rm Higgs}$ have constrained the Higgs boson to have
mass above $114.1\gev$ at the 95\% C.L.~\cite{LEPHWG:2001xw}.  
The remaining $108\gev$ window of possible Higgs mass
is relatively narrow and proposals for future experiments
have focused heavily on this region.  

Nevertheless, a light Higgs boson below $222\gev$ is not guaranteed
for several reasons.  First, precision electroweak data is sensitive
mostly to the logarithm of the Higgs boson mass, and small changes in
the $\chi^2$ fit can yield large changes in the allowed Higgs mass. 
Using the ``all data'' fit from
Table 13.2 of the LEP Electroweak Working Group summary 
report~\cite{Abbaneo:2001ix}, which concludes
that 
\bea
\log_{10}(m_h/{\rm GeV})=1.94^{+0.21}_{-0.22}~~~
({\rm all~data~LEPEWWG~fit}),
\eea
we can deduce that the $3\sigma$ ($4\sigma$) upper bound on the Higgs mass is
about $372\gev$ ($603\gev$). A true value $4\sigma$ away from the 
experimentally
determined central value is by no means out of the question.
Furthermore, it is possible that a much heavier Higgs boson
can conspire with new states ($Z'$ bosons, 
new scalars, etc.) to be compatible with the precision electroweak 
data~\cite{Peskin:2001rw}.
In this article we will not focus on the statistics and solidity
of the present Higgs boson mass limits. Instead, our
main purpose is to determine how much information can be learned
about a theory from the Higgs boson mass whatever it might turn
out to be.

We frame our discussion by first
assuming that a Higgs boson exists that couples to SM states in the
well-defined SM way. We then wish to explore what different values of the
Higgs boson would imply for supersymmetric and non-supersymmetric
gauge unification theories.  Our last ingredient is to take into
consideration an infinite set of possible unification scenarios 
parametrized by the value of $\swsq$ at the unification scale
$\Lambda$.  Most of the previous discussions on Higgs mass limits have
relied strongly on high-energy unification ($\Lambda\sim 10^{16}\gev$).
In our analysis, the scale $\Lambda$ ranges from $\sim 1\tev$ to
$\sim 10^{18}\gev$.

\section{Non-supersymmetric unification}

Many years ago it was discovered that the SM fermions fit very
nicely into two representations of $SU(5)$ and one representation
of $SO(10)$.  Unifying the SM gauge groups also gives explanation
to the unusual values of the hypercharges.  Grand Unification
Theories (GUTs) based on these two groups gained considerable attention 
and are still of value today.

Non-supersymmetric $SU(5)/SO(10)$ GUTs have at least three major challenges.  
They have
no explanation for the huge hierarchy between $M_{\rm GUT}$
and $m_W$. The gauge couplings do not precisely meet at any point
at higher energy.  The precise meeting point of the GUT-normalized 
hypercharge gauge coupling $g_1\equiv g'\sqrt{5/3}$ and the $SU(2)_L$ gauge 
coupling $g_2$ is at a scale that would be too low ($\sim 10^{13}\gev$) 
to satisfy proton
decay constraints if the GUT gauge boson masses were nearby.

Despite all the problems with these high-scale non-supersymmetric
unification scenarios, there have been clever attempts to salvage
them~\cite{minimal GUT}, and probably other clever ways that have
yet to be discussed. 
The various attempts to save them may require intermediate
thresholds, but here we do not admit all those uncertainties and instead
analyze the SM evolution up to the unification scale $M_{\rm GUT}=\Lambda$ 
where
$\swsq=3/8$ ($g_2/g'=\sqrt{5/3}$).
Furthermore, the additional threshold corrections and symmetries 
required to make non-supersymmetric GUTs work may have very little impact
on the Higgs sector.  Or, more likely, there would exist an intermediate
scale $M_I$, perhaps associated with the neutrino seesaw scale,
such that below it one expects perturbative SM evolution.
Therefore, we will keep this idea within the
space of possibilities when discussing Higgs boson predictions, and
we will make plots of Higgs boson mass assuming a perturbative
SM evolution below an arbitrary scale $\Lambda$.  The reader can then
associate $\Lambda$ with $M_{\rm GUT}$, $M_I$, or some other scale
as he or she pleases.

A second idea that motivates unification at the low-energy scale is 
$SU(3)$ electroweak 
unification~\cite{Dimopoulos:2002mv,Li:2002pb,Hall:2002rk,Antoniadis:2000en} 
(for some earlier attempts, 
see~\cite{Weinberg:1971nd,Pisano:ee,Frampton:1992wt}).  
The generic prediction of this framework
is that the hypercharge and $SU(2)_L$ couplings unify at
$\swsq=1/4$ ($g_2/g'=\sqrt{3}$), 
which translates to a scale of about $3.8\tev$.
It is relatively easy within these unification models to adjust
the value of $\swsq$ at the unification scale to values above
$1/4$.  

Our goal now is to demarcate the range of Higgs masses that allow
for a perturbative unification theory, with a perturbative Higgs
self-coupling, for each value of $\swsq$.
We will consider all values of $\swsq$ from $1/4$ to $3/8$.  

The reader should not get the impression that we are advocating theories 
with perturbative couplings as somehow more likely than theories with
non-perturbative couplings.  Nature's reality is probably
independent of the
pain it gives humans to understand it.  Our emphasis here is only that
by analyzing the Higgs sector coupling as given by its mass, we can 
determine if a perturbative theory is compatible with a particular
unification scenario.  One exception to this modest interpretative value
to our work here is supersymmetric GUTs, where it appears desirable
to keep all couplings perturbative so as not to feed into the gauge
coupling renormalization group equations
and spoil the extraordinary unification of all three gauge
couplings of the SM at a high scale.

The Higgs potential for the doublet Higgs field $\Phi$ is
\bea
\label{higgs potential}
V(\Phi)= -m^2_\Phi \Phi^\dagger \Phi +\frac{\lambda}{2}(\Phi^\dagger \Phi)^2.
\eea
After spontaneous symmetry breaking, three of the four degrees of freedom
are eaten by the weak gauge bosons, leaving one physical degree of freedom
remaining with mass,
\bea
m_h^2=\lambda v^2
\eea
where $v^2=1/\sqrt{2}G_F\simeq (246\gev)^2$.

Requiring the theory remain perturbative up to some scale $\Lambda$ implies
that the Higgs self-coupling $\lambda<\lambda_0$, where $\lambda_0$ is
a non-perturbative value for the coupling constant.  Choosing a numerical
value for $\lambda_0$ is not easily justified in this analysis.
One finds many conditions advocated for $\lambda$ to remain perturbative:
$\lambda\lsim {\rm 2}$, $\lambda\lsim \pi$, $\lambda\lsim 4\pi$,
$\beta_\lambda\lsim 1$, etc.  These choices, unfortunately, depend on
the numerical prefactor to the $(\Phi^\dagger\Phi)^2$ coupling.  For
example, if we were to rescale $V\supset 24\lambda(\Phi^\dagger\Phi)^2$ the
conditions on $\lambda$ set forth above are clearly much too weak to
identify the onset of the non-perturbative regime.

We therefore seek a
definition of the onset of non-perturbative behavior which is independent
of the numerical factors out in front of operators.  Our methodology
that satisfies this aim is to turn off all couplings except the
Higgs coupling and expand its $\beta$ 
to successively higher loop order,
\bea
\label{SM Higgs RGE}
\frac{d\lambda}{dt} =  \sum_i \beta^{(i\,{\rm loop})}_\lambda 
= \frac{L_1}{16\pi^2}\lambda^2
  +\frac{L_2}{(16\pi^2)^2}\lambda^3+\frac{L_3}{(16\pi^2)^3}\lambda^4
  +\cdots,
\eea
where in the $\overline{{\rm MS}}$ scheme,
$L_{1,2,3}=12$, $-78$, and $897+504\zeta (3)\simeq 1503$,
respectively~\cite{Jack:sr,Kazakov:1979ik}. 
We then identify the onset of non-perturbativity as when any higher
loop order contribution to the beta function exceeds the value of
any lower loop order contribution.  That is, 
\bea
\label{perturbative condition}
\left|\beta^{(j>i)}_\lambda \right|< 
\left|\beta^{(i)}_\lambda \right|~ ({\rm perturbativity~ condition})
\eea
implies perturbative
coupling, and violation of the condition implies non-perturbative coupling.  
Therefore, given our definition of $\lambda$ from
eqs.~\ref{higgs potential} and \ref{SM Higgs RGE}, $\lambda$ remains
perturbative as long as it is below $\lambda_0=8.2$.

\begin{figure}[bt]
\begin{center}
\includegraphics*[totalheight=3.7in]{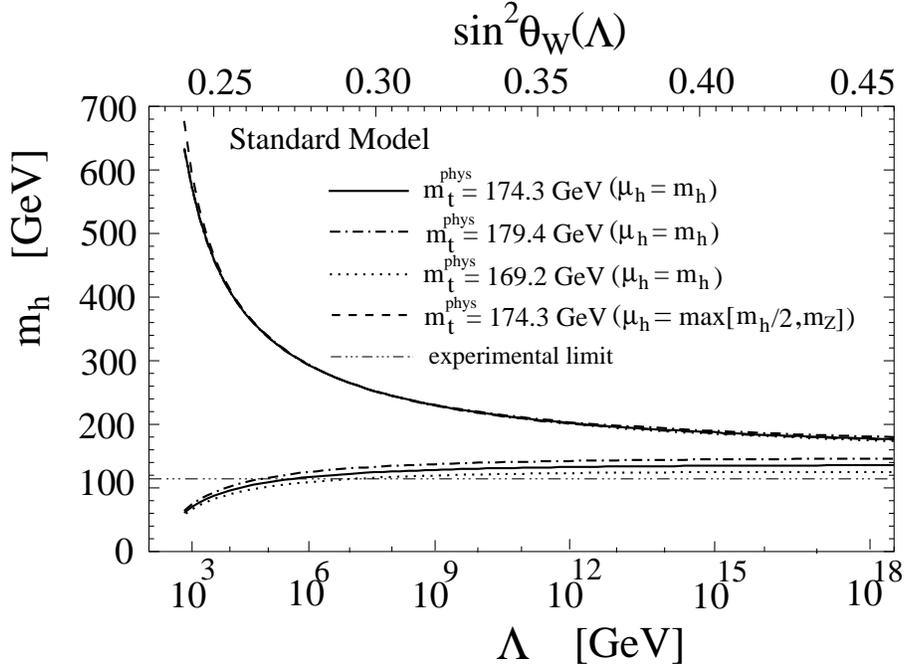}
\caption{The upper curve is the 
limit on the Higgs boson mass within the Standard Model such that
the Higgs self-coupling remains perturbative, i.e., $\lambda<\lambda_0=8.2$,
up the scale $\Lambda$. The lower curve is the Higgs limit such that
$\lambda>0$ for all scales below $\Lambda$. The $x$-axis can be equivalently
expressed as $\Lambda$ or the directly correlated value of
$\swsq(\Lambda)$ which is labeled above.}
\label{mH_limit}
\end{center}
\end{figure}

In fig.~\ref{mH_limit} we plot the upper limit on the Higgs mass that
can be expected in a theory that remains perturbative (i.e., $\lambda<
\lambda_0=8.2$) up to the scale $\Lambda$ (similar SM analyses can
be found in~\cite{Nierste:1995zx}). At
the bottom of the figure we also plot a lower limit
of the Higgs mass by requiring that $\lambda$ remains positive
for all scales below $\Lambda$. This gives an estimate for the
Higgs mass requirement from
vacuum stability~\cite{Altarelli:1994rb}.

The computation of the perturbative limit 
was done using full two-loop renormalization group equations (RGEs)
for the SM gauge couplings, Higgs self-coupling 
and top quark Yukawa coupling. Other couplings are irrelevant to the
analysis of the Higgs mass limit. 
In order to determine the initial condition for RGEs of
gauge couplings at $m_Z$, we adopted experimental values~\cite{Abbaneo:2001ix}
of the QED fine structure constant $\alpha^{-1}=137.06$, the
hadronic contribution to the QED coupling at $m_Z$ 
$\Delta \alpha_{\rm had}^{(5)}(m_Z)=0.02761$, the leptonic
effective electroweak mixing angle $\sin^2\theta_{\rm eff}^{\rm lept}=0.23136$,
and the QCD coupling $\alpha_s(m_Z)=0.118$. We then convert them into
the $\overline{{\rm MS}}$ gauge couplings by using the formula in
refs.~\cite{Fanchiotti:1992tu}.

The one-loop corrections to the  $\overline{{\rm MS}}$ top Yukawa
coupling as a function of the top quark physical mass, $m^{\rm phys}_t$, 
are given
in~\cite{Arason:1991ic,Hempfling:1994ar},
\bea
y_t(\mu)=2^{3/4}G_F^{1/2}m^{\rm phys}_t\left\{ 1+\delta_t(\mu)\right\}.
\eea
We use the full QCD and electroweak contributions to $\delta_t(\mu)$ as
given by eq.~2.12 and the appendix 
of~\cite{Hempfling:1994ar}, and we match it at top quark mass scale
$\mu = m_t^{\rm phys}$.  
The QCD corrections are the dominant 
contribution and have the value,
\bea
\delta^{\rm QCD}_t(\mu)=\frac{\alpha_s}{3\pi}\left[
3\ln \left( \frac{m_t^{{\rm phys}}}{\mu}\right)^2 -4\right].
\eea
We also included all relevant
one-loop finite corrections to set the $\overline{{\rm MS}}$
Higgs coupling as a boundary condition at some scale 
$\mu=\mu_h={\cal{O}}(m_h)$:
\bea
\lambda(\mu) = \sqrt{2} G_F m^2_h \left\{ 1 +\delta_h(\mu)\right\},
\eea
where $\delta_h(\mu)$ is given by eqs.~15a-15d of ref.~\cite{Sirlin:1985ux},
and reproduced in eqs.~B1-B3 of ref.~\cite{Arason:1991ic}.
In fig.~\ref{mH_limit}, taking $\mu_h=m_h$ or max$[m_h/2, m_Z]$, 
we show  the $\mu_h$-dependence on the Higgs mass upper limit.
As the scale $\Lambda$ gets lower, the $\mu_h$-dependence becomes
larger. However, when $\Lambda >3.8\tev$, the difference is
less than $5\gev$ for $\mu_h$ in the range max$[m_h/2, m_Z]<\mu_h<m_h$.

In fig.~\ref{mH_limit}, we also showed a dependence of the top quark mass
on the limits. We used the experimental result 
$m^{\rm phys}_t=174.3\pm 5.1\gev$~(see p.\ 389 of ref.~\cite{topmass}). 
Since the error on the top quark mass is now less than 3\%,
the induced variabilities on the Higgs boson mass
upper limits are almost negligible.

As we run the gauge couplings up to higher scales, the value of $\swsq$
changes. We define $\swsq$ in the $\overline{\rm MS}$ scheme, and
\bea
\swsq(\Lambda)\equiv \frac{{g'}^2(\Lambda)}{{g'}^2(\Lambda)+g_2^2(\Lambda)},
\eea
where $g'(m_Z)\simeq 0.36$ and $g_2(m_Z)\simeq 0.65$.
The correspondence between scale $\Lambda$ and the 
value of $\swsq(\Lambda)$ 
at that scale is plotted on this same graph.  There are
three cases of particular interest in this graph. These are
the perturbative upper limit for $m_h$ when 
$\Lambda=M_{\rm Pl}=2.4\times 10^{18}$ GeV; the
perturbative range for $m_h$ at the scale where $\swsq=3/8$, which
should be close to the unification 
scale of simple $SU(5)$ or $SO(10)$ GUT theories; and, the perturbative
upper limit for $m_h$ at the scale where $\swsq=1/4$, which is
relevant for $SU(3)$ electroweak unification.  We summarize the
results of these three possibilities,
\bea
\nonumber
\swsq=1/4 & \Rightarrow  & \Lambda=3.8\tev, m_h<460 \gev \\
\nonumber
\swsq=3/8 & \Rightarrow & \Lambda\simeq 10^{13}\gev,
m_h<200 \gev \\
\nonumber
\Lambda=M_{Pl} & \Rightarrow  & m_h< 180 \gev .
\eea

When the scale $\Lambda$ is low, we must be concerned that incalculable
non-renormalizable operators might contribute significantly to
the Higgs boson mass. However, even for a low scale such
as $\Lambda=3.8\tev$ needed for $SU(3)$ electroweak unification, the
non-renormalizable operators are not expected to have a large
impact on the Higgs boson mass given our assumptions of perturbativity.
For example, we can look at the simple dimension six operator,
\bea
{\cal{L}}_{NR}=f\frac{|\Phi|^6}{\Lambda^2}
\eea
and estimate the mass shift of the Higgs boson to be
\bea
\Delta m_h^2= f\left[ 31\gev\left( \frac{3.8\tev}{\Lambda}\right)\right]^2.
\eea
Since this adds in quadrature to the Higgs mass, a Higgs mass of
$460\gev$ is increased by only $1\gev$ to $461\gev$ if $f\simeq 1$
and $\Lambda=3.8\tev$, as would be appropriate when considering
the $\swsq(\Lambda)=1/4$ case.


\section{Supersymmetric unification}

\subsection{Minimal Supersymmetric Standard Model}

Within the small uncertainties of perturbative threshold corrections,
all three gauge couplings meet at one point in a simple
grand unified theory if we assume minimal supersymmetric standard
model (MSSM). The unification
scale is about $2\times 10^{16}\gev$. 

Unlike the SM, the Higgs sector in the MSSM
faces very little direct constraint by enforcing perturbativity of
couplings up to the high scale.  Because there is no free
parameter like $\lambda$ in the Higgs potential of minimal supersymmetry,
the only parameters that would be subject to the perturbativity
constraint are the top and bottom Yukawa couplings,
\bea
\label{yukawas mssm}
y_t=\frac{\sqrt{2}\bar m_t(m_Z)}{v\sin\beta},~~~~~
y_b=\frac{\sqrt{2}\bar m_b(m_Z)}{v\cos\beta},~~~~~
y_\tau=\frac{\sqrt{2}\bar m_\tau(m_Z)}{v\cos\beta},
\eea
where $\tan\beta$ is the ratio of the vacuum expectation values of
the two Higgs doublets needed to give mass to the up and down
quarks (see sec.~X of~\cite{Martin:1997ns} for a discussion of the MSSM
Higgs sector), and the $\overline{{\rm DR}}$ masses
$\bar m_t(m_Z)$, $\bar m_b(m_Z)$, and $\bar m_\tau(m_Z)$ are defined
precisely the same as $\hat m_t(m_Z)$, $\hat m_b(m_Z)$, and 
$\hat m_\tau(m_Z)$ in
ref.~\cite{Pierce:1996zz}.   The definitions of the
Yukawa couplings in eq.~\ref{yukawas mssm}
are the same as eq.~(17) of ref.~\cite{Pierce:1996zz}, and the relationship
between the physical masses and the $\overline{{\rm DR}}$ masses
can be found in sec.~3 of ref.~\cite{Pierce:1996zz}.

If $\tan\beta$ is too low (high), the top (bottom)
Yukawa coupling will go non-perturbative before the unification scale.  
Perturbativity up to the high-scale is motivated in this
scenario because non-perturbative couplings would feed into the gauge
coupling RGEs and disrupt the beautiful unification.  Perturbativity
up to this high scale then
puts a constraint on $\tan\beta$ to be within the
range $2\lsim \tan\beta \lsim 65$, which in turn puts a constraint on
the possible values of the lightest Higgs mass in supersymmetry.

We can expand the lightest MSSM Higgs state in terms of a simple, but
useful, equation:
\bea
\label{mssm higgs mass}
m_h^2=m_Z^2\cos^2 2\beta +\eta\frac{3 G_F {m}^4_t}{\sqrt{2} \pi^2}\ln 
\frac{\Delta_S^2}{{m}_t^2}.
\label{MSSM_higgs}
\eea
Here ${m}_t$ denotes the running SM top-quark mass in the 
$\overline{{\rm MS}}$ scheme at the scale $m^{\rm phys}_t$, as utilized
in the Higgs mass computations of ref.~\cite{Carena:2000dp}.
One can think of eq.~\ref{mssm higgs mass} as being valid in the
limit of $m_A\gg m_Z$, or one can absorb the mixing effects between
light and heavy Higgs bosons as being 
absorbed into the $\Delta_S$ definition.
Since it has been known that $O(\alpha \alpha_s)$ two-loop contributions
to the MSSM lightest Higgs mass reduce the one-loop upper limit on 
$m_h$~\cite{Carena:2000dp},
we introduce a suppression factor $\eta$ in eq.~(\ref{MSSM_higgs}). 
To fix $\eta$, we match our expression eq.~(\ref{MSSM_higgs}) with 
the one in
ref.~\cite{Carena:2000dp} at 
$\Delta_S^2=m^2_{\tilde t}\equiv (m_{\tilde t_1}^2
+m_{\tilde t_2}^2)/2
=(1\tev)^2$ assuming no stop mixing, and then we get
\bea
\eta=1-\frac{2\alpha_s}{\pi}\left(
\ln \frac{m_{\tilde t}^2}{{m}_t^2}-\frac{2}{3} \right)
=0.78.
\eea
The numerical value of $\Delta_S$ is therefore a good
indicator of the scale of superpartner masses.

Also, we remark that
the quantity $\Delta_S$ has been introduced in analogy 
to $T_{\rm SUSY}$ in refs.~\cite{Langacker:1992rq,Carena:1993ag}. 
$T_{\rm SUSY}$ is useful because
it remaps all superpartner threshold effects into one single mass 
scale for the purposes of matching gauge couplings
between the SM effective theory below $T_{\rm SUSY}$ to the
fully supersymmetric theory above $T_{\rm SUSY}$.  The purpose
of $\Delta_S$ is similar.  It can be defined as the matching scale 
that reproduces the correct Higgs mass by
running supersymmetric RGEs above it and the one-loop
SM RGE for $\lambda$ below it, assuming (correctly) that the $y_t^4$ part
of $\beta_\lambda$ dominates.

\begin{figure}[bt]
\begin{center}
\includegraphics*[totalheight=3.7in]{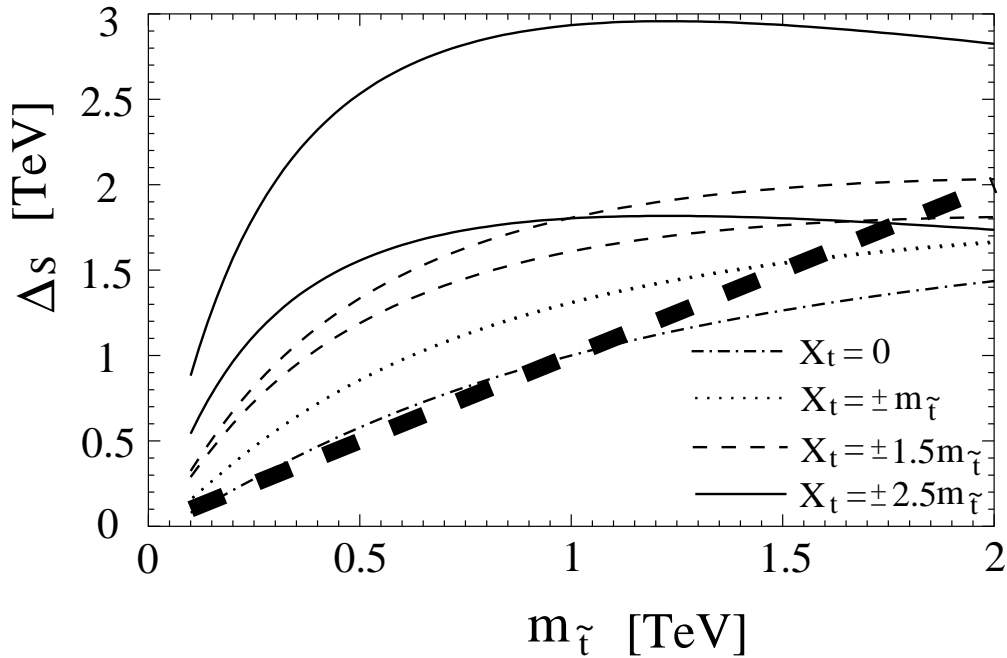}
\caption{The relationship between $\Delta_S$ defined by 
eq.~\ref{mssm higgs mass} and 
$m_{\tilde t} \equiv \sqrt{(m^2_{\tilde t_1}+m^2_{\tilde t_2})/2}$
for various top squark mixing $X_t=A_t-\mu\cot\beta$ in 
the limit $m_A\gg m_Z$ (no Higgs mixing effects).  Since
$\Delta_S>m_{\tilde t}$ for much of parameter space, superpartners
are expected to be below the value of $\Delta_S$ that corresponds
to the Higgs boson mass limit $m_h>114.1\gev$. For the reader's convenience
a thick dashed line is plotted for the line $\Delta_S=m_{\tilde t}$.}
\label{stops and dels}
\end{center}
\end{figure}

\begin{figure}[tb]
\begin{center}
\includegraphics*[totalheight=3.7in]{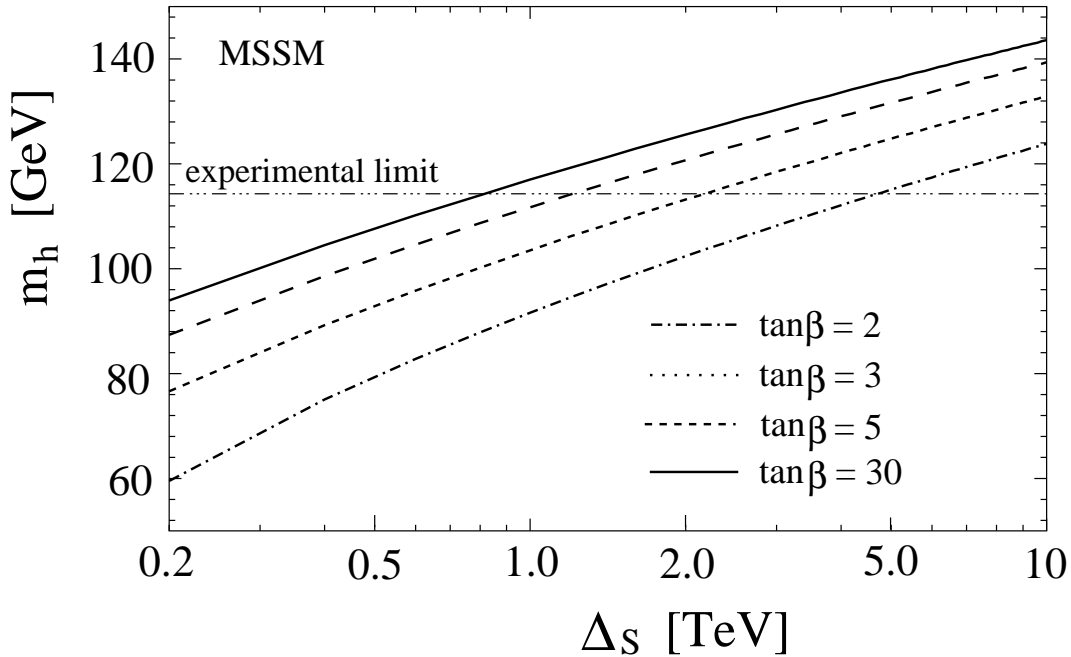}
\end{center}
\caption{The lines plot the lightest Higgs boson mass in
the MSSM as a function of the supersymmetry scale $\Delta_S$,
whose leading log value is $\Delta_S^2=m_{\tilde t}^2$.
The four lines from bottom to top represent $\tan\beta = 2,3,5,30$.}
\label{mssm_higgs}
\end{figure}

Top squark mixing effects will begin to decorrelate the value of
stop masses from that of the correct value of
$\Delta_S$ needed to recover an accurate Higgs mass using
eq.~\ref{mssm higgs mass}.  For fixed stop masses, the higher the
left-right mixing effects the larger the Higgs mass becomes, and therefore
the larger $\Delta_S$.  We demonstrate this effect in fig.~\ref{stops and dels}
by computing the needed value of $\Delta_S$ to reproduce the correct
Higgs mass given various values of the mixing term in the stop mixing
matrix $X_t=A_t-\mu\cot\beta$ (see eq.~(4) of~\cite{Carena:2000dp}).
We have computed the Higgs mass using eqs.~(46) and (47) 
of~\cite{Carena:2000dp}, and then recast the Higgs mass in the equivalent
variable $\Delta_S$ using eq.~\ref{mssm higgs mass}.  For various values
of the stop mixing $X_t$ we have plotted the correlation between
required $\Delta_S$ 
and $m_{\tilde t}$.
As we see, $\Delta_S>m_{\tilde t}$ for most of parameter space, which
enables us to conclude that $\Delta_S$ generally
overestimates the supersymmetry mass scales of the top squarks in
the presence of mixing.  This should be kept in mind when interpreting
the parameter space for superpartners from fig.~\ref{mssm_higgs}; that is,
superpartners can be significantly lighter than the $\Delta_S$ values
needed for a Higgs mass above the experimental limit.

In fig.~\ref{mssm_higgs} we have plotted the lightest Higgs mass
in the MSSM as a function of the supersymmetry mass scale $\Delta_S$.
Although presented in a slightly different way here, the results of
this plot are well known~\cite{Okada:1990vk}.  
Low $\tan\beta$ requires large supersymmetry
breaking mass in order to evade the current experimental limits on
the lightest Higgs boson mass.  Large $\tan\beta$ enables the MSSM
to be comfortably within all experimental constraints
for moderately small supersymmetry breaking mass.  

We view it as a success that supersymmetry predicts that its
lightest Higgs boson mass can naturally reside in the squeezed window
of $114.1\gev$ from direct experimental searches and $222\gev$ from
precision electroweak measurements. Superpartners could disrupt the
precision electroweak predictions, but it is well known that supersymmetry
decouples rapidly from $Z$-pole observables. Attempts to make the
global fits to the data better by resolving some small discrepancies
between leptonic and hadronic observables also demonstrate the 
rapid decoupling of supersymmetry, since the active superpartners in
these studies must be very light (e.g., 
$m_{\tilde \nu}\lsim m_Z$ as in ref.~\cite{Altarelli:2001wx}).

\subsection{The next-to-minimal supersymmetric standard model}

As soon as one goes beyond the most minimal supersymmetric theory,
the constraints of perturbativity become very significant again, just
as they were in our SM 
analysis~\cite{Espinosa:1991gr,Kane:1992kq,Espinosa:1998re,Masip:1998jc}. 
The reason is because non-minimal
supersymmetric theories add additional Yukawa couplings that contribute
directly to the mass of the lightest Higgs, but are not usefully constrained
by any known measurement.

The most important example of non-minimal supersymmetry is
the NMSSM (next-to-minimal MSSM), 
which adds another singlet $S$ to the theory.  This
approach has been used by many authors to make the $\mu$ term
more natural within supersymmetry. That is, in the MSSM there exists
a term in the superpotential $\mu H_uH_d$ which might be best explained
by an NMSSM term, $\lambda_s SH_uH_d$, where 
$\mu=\lambda_s\langle S\rangle$.

We can write the mass of the lightest scalar of the NMSSM theory in
a very similar way as we did for the MSSM:
\bea
\label{nmssm higgs}
m_h^2 = m_Z^2\cos^2 2\beta +\frac{\lambda^2_s}{2\sqrt{2}G_F}\sin^2 2\beta 
+\eta \frac{3 G_F m^4_t}{\sqrt{2}\pi^2}\ln \frac{\Delta_S^2}{m_t^2},
\eea
where we take $\eta=0.78$.
The scale $\Delta_S$ is close but not precisely the same as it is in the MSSM.
This is because there are some more states and parameters in the NMSSM
that feed into $\Delta_S$, such as the additional masses and mixings in
the Higgs sector. There could also be additional contributions to the
lightest mass, and to $\Delta_S$, if $S$ is charged under another gauge
group.  In that case, $H_u$ and/or $H_d$ would be charged too, leading
to additional contributions to the mass~\cite{Drees:tp}.  
Furthermore, a large Yukawa coupling of a fourth generation 
to the Higgs boson can add substantial radiative corrections
to the Higgs mass, just as the top Yukawa does in the 
MSSM~\cite{Moroi:1992zk}. 
For the purposes
of being conservative and illustrative of how even the smallest
deviation from the MSSM can affect the lightest Higgs mass, 
we will ignore additional gauge charges or additional states that
may contribute to the radiative corrections of the Higgs mass.

The numerical value of $\lambda_s$ is arbitrary.  If it is large it
contributes significantly to the Higgs mass via eq.~\ref{nmssm higgs}
and raises it to a much higher value than the MSSM prediction for
the same values of $\tan\beta$ and $\Delta_S$.  However, if we do not
wish to spoil perturbative gauge coupling unification we must require
that $\lambda_s$ and the other remaining couplings, such as $y_t$ and
$y_b$, remain perturbative so as not to disrupt too much the RGE evolution
of the gauge couplings.  

The one-loop $\beta$ functions of
$\lambda_s$, $y_t$ and $y_b$ all depend on each other.  Therefore, we
must insure that all remain perturbative.  To determine what values
of $\lambda_s$, $y_t$ and $y_b$ are perturbative, we employ
the condition of eq.~\ref{perturbative condition} on each of these
three couplings.  To do this we compute the three-loop $\beta$ functions
using refs.~\cite{Jack:1996qq}
for each of the couplings in the limit that all other couplings are
turned off:
\bea
\beta_{y_t}^{(3\, {\rm loop})} 
& = & \frac{6}{16\pi^2}y_t^3-\frac{22}{(16\pi^2)^2}y_t^5
 +\frac{(102+36\zeta(3))}{(16\pi^2)^3}y_t^7 \\
\beta_{\lambda_s}^{(3\, {\rm loop})}& = & \frac{4}{16\pi^2}\lambda_s^3
  -\frac{10}{(16\pi^2)^2}\lambda_s^5
 +\frac{(32+24\zeta(3))}{(16\pi^2)^3}\lambda_s^7 .
\eea
$\beta_{y_b}$ is the same as $\beta_{y_t}$ after replacing $y_t\to y_b$.
Applying the perturbativity conditions of
eq.~\ref{perturbative condition} we find that perturbative
couplings must satisfy $y_t<4.9$, $y_b<4.9$, 
and $\lambda_s<5.1$.

\begin{figure}[tb]
\begin{center}
\includegraphics[totalheight=3.7in]{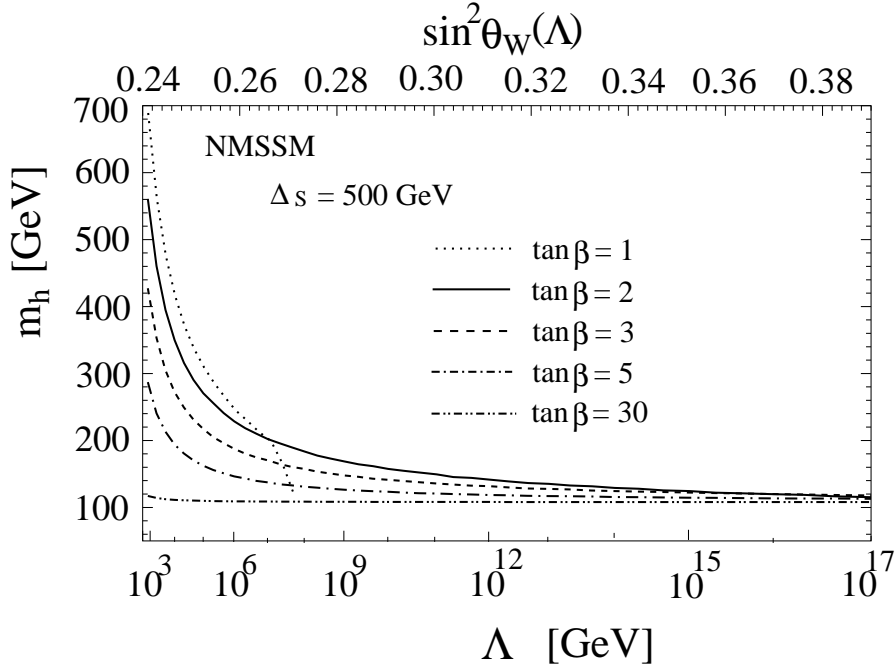}
\end{center}
\caption{The five lines are for the same values of $\tan\beta$ in the
NMSSM.  The $\lambda_s$ coupling of the superpotential $\lambda_s SH_uH_d$ term
is assumed to be at its maximum allowed value without blowing up
before the scale $\Lambda$ ($\lambda_s<5.1$). 
Since $\swsq(\Lambda)$ correlates directly
with $\Lambda$ we provide the $\swsq(\Lambda)$ values on the upper axis.}
\label{nmssm}
\end{figure}

In fig.~\ref{nmssm} we plot the mass of the lightest Higgs boson as
a function of scale $\Lambda$, requiring that all couplings remain
perturbative below $\Lambda$.  
In this analysis, we use two-loop RGEs for all gauge, top and bottom Yukawa
and Higgs couplings
including full one-loop supersymmetry corrections to the Yukawa couplings
discussed below eq.~\ref{yukawas mssm},
and one-loop supersymmetry logarithmic corrections to all gauge couplings
at $m_Z$. For this computation we set all supersymmetry masses
to $\Delta_S$.
To be consistent with our definition
of $\Delta_S$ given above, the scale at which $\lambda_s$ is
evaluated in eq.~\ref{nmssm higgs} is $\Delta_S$.
The five different curves in the
figure represent different values of $\tan\beta=1$, 2, 3, 5, and 30.
In the MSSM (without GUT), $\tan\beta<1$ is excluded by the Higgs search.
However, in the NMSSM with low $\Lambda$, such low values of $\tan\beta$
are allowed as long as constraints such as $b\to s\gamma$,
for example, are satisfied, which would perhaps require a very
heavy charged Higgs mass (see, e.g., fig.~12 of ref.~\cite{Hewett:1996ct}).

We can then look again at the most interesting scales in this plot
related to $\swsq=1/4$ $SU(3)$ electroweak unification, and $\swsq=3/8$
$SU(5)/SO(10)$ grand unification.  In the NMSSM the $\Lambda$ scales
associated with this unification are different than they were in
the SM case.  Furthermore, the unknown Yukawa coupling $\lambda_s$ 
contributes to the mass of the lightest Higgs boson in a very different
way than the SM Higgs self-coupling $\lambda$, and the RGEs are 
very different.  The result, with $\Delta_S=500\gev$ is
\bea
\nonumber
\swsq=1/4 & \Longrightarrow & \Lambda=37\tev, m_h<350 \gev, \\
\nonumber
\swsq=3/8 & \Longrightarrow & \Lambda\simeq 2\times 10^{16}\gev,
m_h< 120 \gev.
\eea

\begin{figure}[tb]
\begin{center}
\includegraphics[totalheight=3.7in]{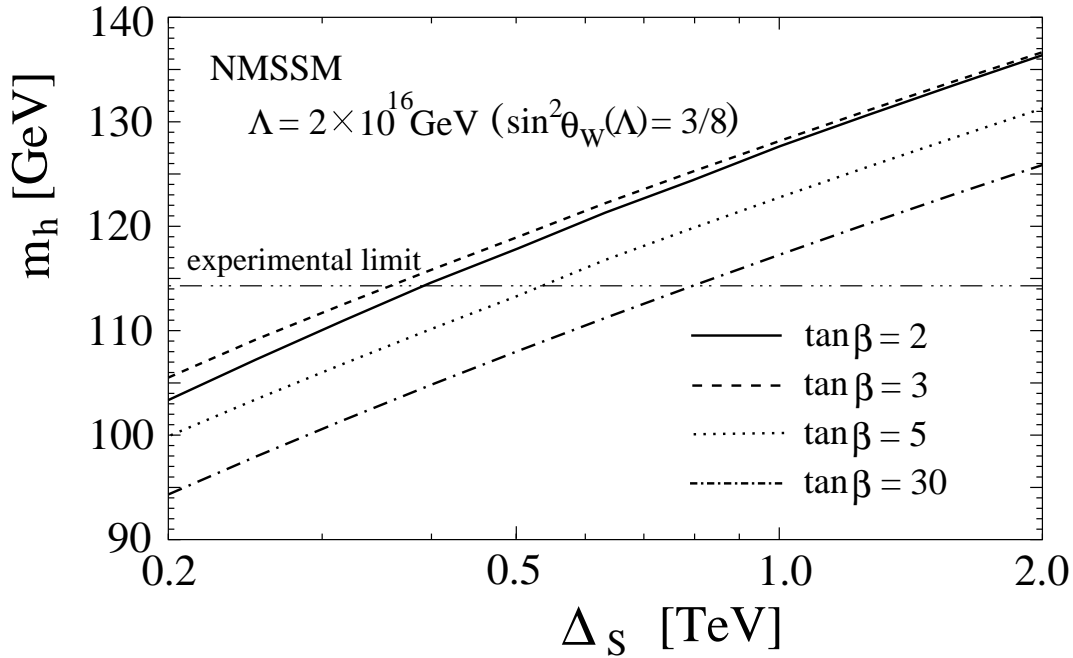}
\caption{The lines plot the lightest Higgs boson mass in
the NMSSM as a function of the supersymmetry scale $\Delta_S$.
The leading log value for $\Delta_S=m_{\tilde t}$.
The value of $\lambda_s$ used in eq.~\ref{nmssm higgs} is at its maximum
consistent with $\lambda_s(\Lambda)<5.1$ (perturbative).  
Here $\Lambda=2\times 10^{16}\gev$,
which corresponds to the simple grand unification scenario
of $\swsq(\Lambda)=3/8$.}
\label{nmssm1}
\end{center}
\end{figure}

\begin{figure}[tb]
\begin{center}
\includegraphics[totalheight=3.7in]{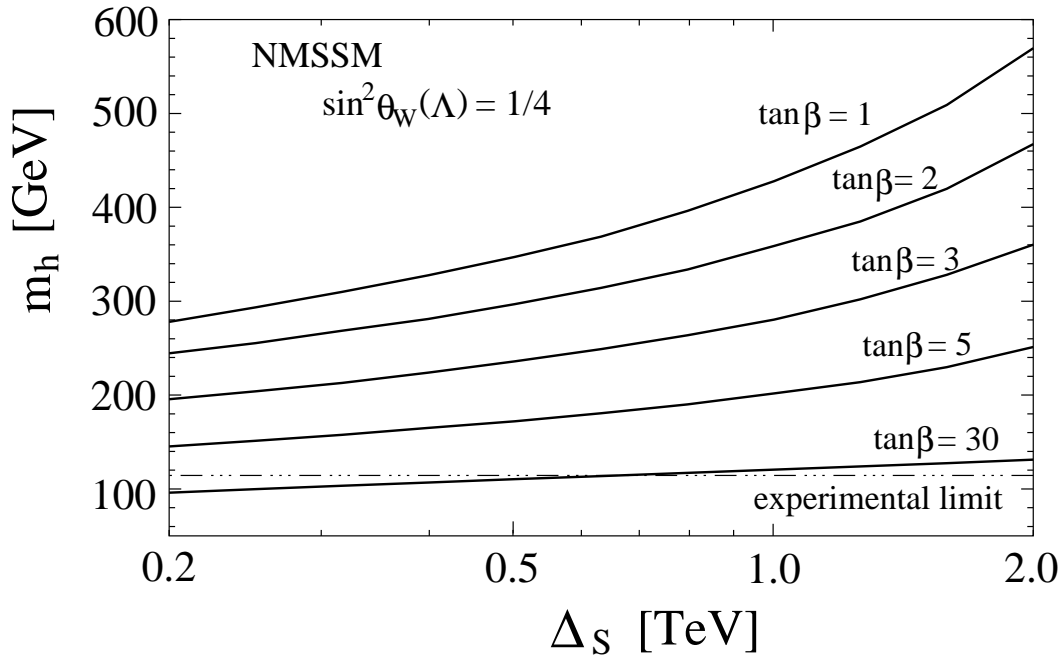}
\caption{The lines plot the lightest Higgs boson mass in
the NMSSM as a function of the supersymmetry scale $\Delta_S$.
The leading log value for $\Delta_S=m_{\tilde t}$.
The value of $\lambda_s$ used in eq.~\ref{nmssm higgs} is such that
$\lambda_s(\Lambda)<5.1$ (perturbative).  Here $8\tev < \Lambda< 110\tev$
(precise value depends on $\Delta_S$),
which corresponds to the $SU(3)$ electroweak unification scenario
of $\swsq(\Lambda)=1/4$.}
\label{nmssm2}
\end{center}
\end{figure}

We plot in figs.~\ref{nmssm1} and~\ref{nmssm2}
the lightest Higgs mass in the NMSSM as a function of 
$\Delta_S$ confining ourselves to the two scenarios $\swsq=1/4$ 
($\Lambda\sim 8-110\tev$) and $\swsq=3/8$ ($\Lambda=2\times 10^{16}\gev$).
For low values of $\tan\beta$ the Higgs mass prediction is very well
separated between the two theories because the $\lambda_s$ contribution
is not suppressed much by $\sin^2 2\beta$ and the difference between
the $\lambda_s(\Delta_S)$ allowed such that $\lambda_s$ is still perturbative
at $\Lambda$ is dramatically different for $\Lambda \sim 10\tev$ 
($\lambda_s(\Delta_S=500 \gev)\sim 2$ allowed) and
$\Lambda=2\times 10^{16}\gev$ ($\lambda_s(\Delta_S=500 \gev)\sim 0.7$ allowed).
However, as we go to higher values of $\tan\beta$ the Higgs mass has
very little dependence on the $\lambda_s^2\sin^2 2\beta$ term since
it is suppressed by $1/\tan\beta$ at high $\tan\beta$.  For that reason,
the two $\tan\beta=30$ lines are very nearly on top of each other
in the plot.

\section{Conclusions}

In this article we have examined Higgs mass upper limits in theories
that are perturbative up to a scale $\Lambda$.  After discovery of
a Higgs boson,
the results of these computations can tell us at what scale a perturbative
description of our low-scale theory (SM, MSSM, NMSSM, etc.) breaks down.
The results can also help determine if unification with $\swsq(\Lambda)=1/4$
or $\swsq(\Lambda)=3/8$ can occur perturbatively.  In the several unification
scenarios we studied,
we found that it is not expected to have a Higgs boson above about
$500\gev$ and still remain perturbative.
This is not a theorem, but a highly suggestive result based on the simple
theories that are currently attractive.

Fortunately, the LHC will be able to see all
SM-like Higgs bosons easily up to $500\gev$ and probably to at least
as high as $800\gev$~\cite{LHC:TDR}.  The extra dynamics that go along
with the explanation for electroweak symmetry breaking, such as supersymmetry
or extra dimensions, should also be detectable at the LHC.  Finding an
NMSSM Higgs boson in some regions of parameter space could be a significant
challenge at the LHC, but there are good indications that the LHC will
cover those possibilities also~\cite{Ellwanger:2001iw}.

A future
linear collider will be able to see and carefully study a Higgs with mass
$m_h\lsim \sqrt{s}-m_Z$ in $e^+e^-$ mode, and perhaps slightly higher
in $\gamma\gamma$ mode~\cite{TESLA,Abe:2001wn,Asner:2001ia}.  
Indications of $SU(3)$ electroweak unification could also come from
direct collider probes and precision electroweak studies
that match 
expectations~\cite{Csaki:2002bz} of minimal models and beyond. 
Directly confirming 
the unification scenarios may be difficult to do, but 
additional clues from measurements of 
superpartner masses and the complete Higgs sector, 
for example, would be critical information if we are to be successful.


\bigskip

\noindent
{\it Acknowledgments: }
KT and JDW were supported in part by the U.S.~Department of
Energy and the Alfred P. Sloan Foundation.

\def\Journal#1#2#3#4{{#1} {\bf #2}, #3 (#4)}
\def\add#1#2#3{{\bf #1}, #2 (#3)}

\def\NPB{{\em Nucl. Phys.} B}
\def\PLB{{\em Phys. Lett.}  B}
\def\PRL{{\em Phys. Rev. Lett.}}
\def\PRD{{\em Phys. Rev.} D}
\def\PR{{\em Phys. Rev.}}
\def\ZPC{{\em Z. Phys.} C}
\def\SJNP{{\em Sov. J. Nucl. Phys.}}
\def\AnnP{{\em Ann. Phys.}}
\def\JETPL{{\em JETP Lett.}}
\def\LMP{{\em Lett. Math. Phys.}}
\def\CMP{{\em Comm. Math. Phys.}}
\def\PTP{{\em Prog. Theor. Phys.}}


\end{document}